# Effective Field Model of Roughness in Magnetic Structures

Serban Lepadatu

*Jeremiah Horrocks Institute for Mathematics, Physics and Astronomy, University of Central Lancashire, Preston PR1 2HE, U.K.*

An effective field model is introduced here within the micromagnetics formulation, to study roughness in magnetic structures, by considering sub-exchange length roughness levels as a perturbation on a smooth structure. This allows the roughness contribution to be separated, which is found to give rise to an effective configurational anisotropy for both edge and surface roughness, and accurately model its effects with fine control over the roughness depth without the explicit need to refine the computational cellsize to accommodate the roughness profile. The model is validated by comparisons with directly roughened structures for a series of magnetization switching and domain wall mobility simulations and found to be in excellent agreement for roughness levels up to the exchange length. The model is further applied to vortex domain wall mobility simulations with surface roughness, which is shown to significantly modify domain wall movement and result in dynamic pinning and stochastic creep effects.

# I. INTRODUCTION

It is well known that roughness in magnetic structures significantly modifies the physical behaviour of magnetic systems, including changes in coercivity, pinning of domain walls for both field and current-driven regimes and resonance phenomena [1-4] Physical roughness is typically modelled by removing cells in a finite difference mesh [5-7] or deforming the mesh for finite element methods [8]. Other methods used to model defects and roughness rely on varying the saturation magnetization [9], changing the exchange stiffness constant at grain boundaries [10] or introducing pinning potentials in collective coordinate models [11,12]. For small roughness levels the usual rough mesh method becomes problematic to simulate as very small cellsize values are required to produce roughness profiles. For surface roughness studies, where roughness levels of the order 1 nm or smaller are typical [13,14], full micromagnetics simulations using roughness profiles physically mapped onto the mesh are very difficult to implement, in particular for finite difference schemes. This work introduces a new method to accurately model small roughness levels, both edge and surface roughness, without the requirement to refine the mesh specifically to accommodate a roughness profile, by introducing a new energy term. Roughness in magnetic films is known to modify surface anisotropy [15] as well as induce a configurational anisotropy by modifying the magnetostatic energy [16,17]. The possibility of treating roughness within micromagnetics as a perturbation on a smooth magnetic body is investigated here. The model introduced has some similarities to the stair-step correction method [18] although it starts from a different approach, has different aims, and the working equations are different. By concentrating on roughened structures, this model separates a roughness energy contribution which can be treated as a perturbation on a smooth structure, and analysed as a separate term, allowing for elegant physical interpretation of results.

In the following sections the model is first introduced and the roughness energy density terms are defined. This is tested by a series of comparisons with the standard rough mesh method, including magnetization switching and domain wall mobility calculations. A discussion of the dependence of the roughness energy density terms on dimensions and roughness depth is given and finally the model is applied to domain wall mobility calculations with surface roughness.

## II. EFFECTIVE FIELD MODEL

The magnetostatic field in a magnetic material is obtained as a convolution between the magnetization vector function ***M***, and the demagnetizing tensor ***N***, Equation 1, where $V$ denotes the magnetic body. [19]

$$\mathbf{H}_d(\mathbf{r}_0) = -\int_{\mathbf{r} \in V} \mathbf{N}(\mathbf{r} - \mathbf{r}_0) \otimes \mathbf{M}(\mathbf{r}) d\mathbf{r} \tag{1}$$

The demagnetizing tensor has 3 diagonal elements, denoted $N_{xx}$, $N_{yy}$, $N_{zz}$ and 3 distinct off-diagonal elements, $N_{xy} = N_{yx}$, $N_{xz} = N_{zx}$ and $N_{yz} = N_{zy}$, which may be calculated using the formulas given by Newell et al. [20]. For a uniformly magnetized magnetic body $V$, it can easily be shown that the magnetostatic energy density term along the longitudinal (*X*-axis) direction of the body is given by ($M_S$ is the saturation magnetization):

$$\varepsilon_{SL}(\mathbf{r}_0) = \frac{\mu_0 M_S^2}{2} \sum_{\mathbf{r} \in V} N_{xx}(\mathbf{r} - \mathbf{r}_0) \quad (\mathbf{r}_0 \in V) \tag{2}$$

Similar equations hold for the transverse (Y-axis) and perpendicular (Z-axis) energy density terms, $\varepsilon_{ST}$ and $\varepsilon_{SP}$, by replacing $N_{xx}$ with $N_{yy}$, or $N_{zz}$ respectively, throughout, thus in what follows only the longitudinal term is shown explicitly for brevity.

Now consider the same magnetic body but with edge and/or surface roughness added, which we will denote $Vr$. The longitudinal, transverse and perpendicular energy density terms for the roughened body, $\varepsilon_{RL}$, $\varepsilon_{RT}$ and $\varepsilon_{RP}$, are obtained as for Equation 2. We introduce the roughness energy density terms, $\varepsilon_L$, $\varepsilon_T$ and $\varepsilon_P$, as additional energy density terms superimposed on the smooth body $V$, such that $<\varepsilon_L>_V = <\varepsilon_{RL}>_{Vr} - <\varepsilon_{SL}>_V$, where $< >_V$ denotes averaging over the magnetic body $V$. Thus we obtain the following expression ($N_V$ and $N_{Vr}$ are the number of cells in the smooth and roughened magnetic bodies, counted on the same mesh, respectively):

$$\langle \varepsilon_L \rangle_V = \frac{1}{N_V} \sum_{\mathbf{r}_0 \in V} \frac{\mu_0 M_S^2}{2} \sum_{\mathbf{r} \in V} N_{xx}(\mathbf{r} - \mathbf{r}_0) G(\mathbf{r}, \mathbf{r}_0)$$

$$G(\mathbf{r}, \mathbf{r}_0) = \begin{cases} \dfrac{N_V}{N_{Vr}} - 1 & \mathbf{r} \wedge \mathbf{r}_0 \in Vr \\ -1 & \mathbf{r} \vee \mathbf{r}_0 \in V - Vr \end{cases}$$

(3)

In light of Equation 3 we can now introduce the roughness energy density function as:

$$\varepsilon_L(\mathbf{r}_0) = \frac{\mu_0 M_S^2}{2} \sum_{\mathbf{r} \in V} N_{xx}(\mathbf{r} - \mathbf{r}_0) G(\mathbf{r}, \mathbf{r}_0) \quad (\mathbf{r}_0 \in V) \tag{4}$$

For a uniformly magnetized body along any direction, similarly we obtain the following expression ($m_x$, $m_y$ and $m_z$ are the normalized magnetization components):

$$\varepsilon(\mathbf{r}_0) = \varepsilon_L(\mathbf{r}_0) m_x^2 + \varepsilon_T(\mathbf{r}_0) m_y^2 + \varepsilon_P(\mathbf{r}_0) m_z^2 \quad (\mathbf{r}_0 \in V) \tag{5}$$

In Equation 5, the off-diagonal demagnetizing tensor elements, involving cross-products of magnetization components, may safely be omitted since at all points and for all magnetization directions they are many orders of magnitude smaller than their diagonal counterparts (also see Figure 1(a) for verification of Equation 5). For edge roughness, the magnetostatic energy density, averaged over the magnetic body, along the perpendicular direction is identical for the smooth and roughened bodies since the roughness depth is uniform along the perpendicular direction, thus $\varepsilon_P$ is zero in Equation 5; for surface roughness this is no longer the case and all energy density terms must be used.

So far we have only considered uniformly magnetized bodies. For non-uniform, but smoothly varying, magnetization configuration with small roughness levels [21], Equation 5 also applies. The following approximation in obtaining Equation 5 is used in this case:

$$N_{xx}(\mathbf{r}-\mathbf{r}_0)M_x(\mathbf{r})M_x(\mathbf{r}_0) \cong N_{xx}(\mathbf{r}-\mathbf{r}_0)M_x^2(\mathbf{r}_0) \tag{6}$$

The verification of Equation 6 is partly the purpose of validating the effective field model against the standard rough mesh method in the following sections. The reasons Equation 6 should hold are, on the one hand the magnetization varies slowly and since the values of the demagnetizing coefficients drop very quickly as $|r\text{-}r_0|$ increases, then $N_{xx}(\mathbf{r}-\mathbf{r}_0)[M_x(\mathbf{r}) - M_x(\mathbf{r}_0)]$ tends to zero very rapidly. Moreover, for small roughness levels, since $N_V/N_{Vr} \approx 1$, the largest contributions in Equation 4 involve terms for which $r \vee r_0 \in V\text{-}Vr$. For small roughness levels these points tend to be spaced relatively far apart, improving the approximation in Equation 6 even further. Thus we may consider Equation 5 as generally valid when implementing the effective field roughness model. This model may now be implemented by pre-calculating the energy terms using Equation 4 for the entire mesh, for a given roughness profile, and deriving the effective roughness field from Equation

5. [22] The roughness field involves minimal computation at run-time since no inter-cell interactions need be included; this field is similar to an anisotropy contribution and is given by:

$$\mathbf{H}_R(\mathbf{r}_0) = -\left[\sum_{\mathbf{r}\in V}\mathbf{N}_{diag.}(\mathbf{r}-\mathbf{r}_0)G(\mathbf{r},\mathbf{r}_0)\right]\mathbf{M}(\mathbf{r}_0) \quad (\mathbf{r}_0 \in V) \tag{7}$$

When choosing the computational cellsize it is important to keep the change in magnetization angle from one cell to another small, since on the micromagnetics length scale the magnetization is formulated as a continuous function. A good rule is to set the cellsize small enough so that further refinement does not produce different results. An important length scale is the exchange length, Equation 8 where $A$ is the exchange stiffness, taking into account the competition between direct exchange coupling and magnetostatic interaction. [23]

$$l_{ex} = \sqrt{\frac{2A}{\mu_0 M_S^2}} \tag{8}$$

Typically the cellsize should be smaller or equal to the exchange length, although for magnetization configurations involving rapid changes in magnetization, such as cross-tie structures [24], cellsize values down to half the exchange length are required. In this work roughness depth values up to the exchange length are considered. For calculations where the roughness profile is included explicitly the cellsize must be chosen small enough so that the details of the profile are reproduced. For the effective field model however, the cellsize is chosen as for the smooth structure, i.e. with enough detail to reproduce changes in magnetization accurately, and the roughness contribution is included separately using the effective roughness field in Equation 7. Validation tests in the following sections will reveal that the effects of a small roughness level profile varying quicker than the coarse cellsize can be accurately reproduced through the effective field terms in Equation 7 at the coarse cellsize. In other words it is not necessary to specifically reduce the cellsize, beyond what is required for continuum approximation, in order to reproduce the effects of a roughness profile.

## A. Roughness Energy Density Configuration

Equation 5 is easily verified by calculating the roughness energy density using the standard rough mesh method directly, then using the effective field method (Equation 5). A typical roughness energy density plot is shown in Figure 1 for a uniformly magnetized 320×80×5 nm $Ni_{80}Fe_{20}$ rectangular prism with 2.5 nm edge roughness depth on both longitudinal edges. Granular roughness profiles were used here with mean grain diameter equal to the roughness depth, however the model is generally applicable to any type of roughness profile. Details of the computational methods are given in Appendix A. For the mesh method a 1.25 nm cellsize was used, whilst for the effective field method a 5 nm cellsize was used. In all the work that follows, all prisms, or wires, lie in the XY plane, are elongated along the X axis (longitudinal direction), edge roughness is applied to both edges along the longitudinal direction and surface roughness to the top surface only. Figure 1(a) shows the sub-exchange length roughness level results in a type of configurational anisotropy with easy axis oriented in the transverse direction. The excellent agreement between the two methods confirms the validity of Equation 5 for uniform magnetization. The model introduced here only considers changes in magnetostatic energy when roughness is introduced, without changes to the exchange energy. This is justified since the magnetostatic energy is generally much larger – validation tests in the next section also show that this is a good approximation. Changes to other terms which do not involve inter-cell interactions, such as magneto-crystalline anisotropy and Zeeman energy, are even smaller since $N_V \approx N_{Vr}$.

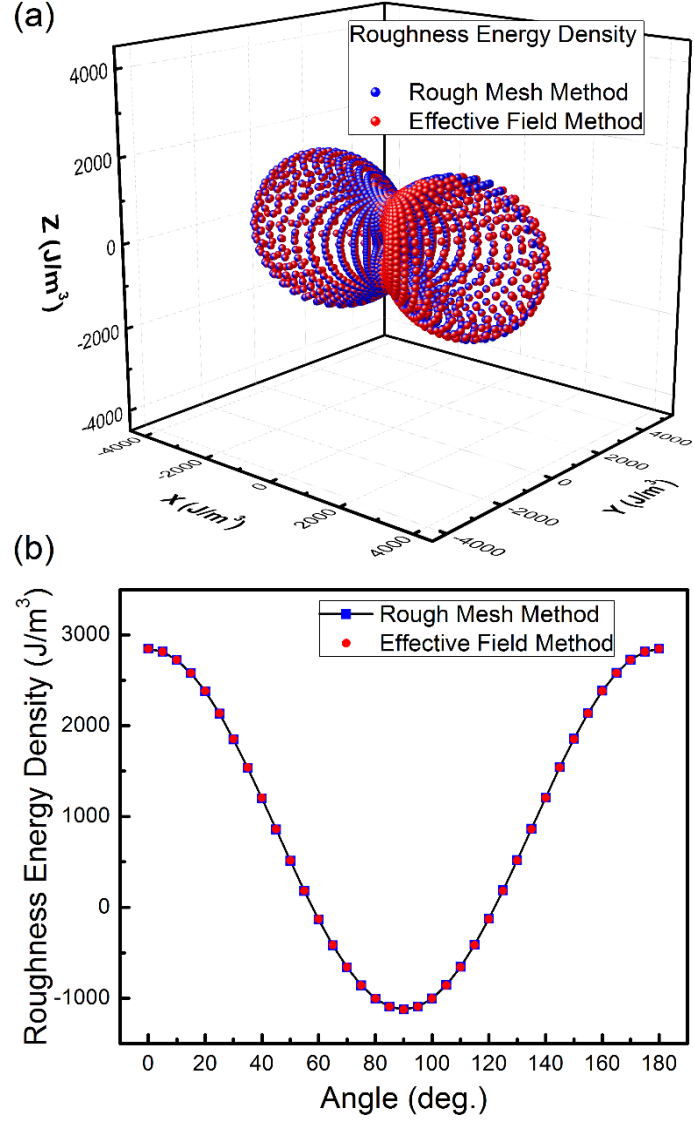

FIG. 1: (color online). Roughness energy density calculated for a uniformly magnetized 320×80×5 nm prism with 2.5 nm roughness depth in the XY plane and oriented along the X axis, using both the rough mesh and effective field methods. (a) Full 3D roughness energy density configuration with the lowest energy point shifted to zero and (b) XY-plane cross-section in the raw energy profile showing the $\varepsilon_L$ (0 deg.) and $\varepsilon_T$ (90 deg.) terms.

## III. VALIDATION

### A. Magnetization Switching

To investigate the validity of the effective field roughness model for non-uniform magnetization, first magnetization switching simulations for a set of rectangular $Ni_{80}Fe_{20}$ prisms with varying roughness levels are described. The magnetization was saturated along the longitudinal direction and the field increased until magnetization reversal occurred. Granular edge roughness profiles were generated with different roughness levels from 1.25 nm up to 5 nm. The standard rough mesh method was used to simulate the magnetization switching events using a cellsize of 1.25 nm. These simulations were then repeated using the effective field roughness method superimposed on a smooth prism using a cellsize of 5 nm. Tests using cellsize values of 2.5 nm showed the same results. This shows it is sufficient to consider an average roughness energy density, or roughness effective field, from a roughness profile varying quicker than the exchange length, at a coarse cellsize where the magnetization varies slowly enough for a good continuum approximation. Typical energy density (sum of magnetostatic and exchange energy density) plots as a function of applied magnetic field are shown in Figure 2(a) – the two methods are in excellent agreement over this range of roughness depth. The coercive field obtained as a function of roughness depth for a set of prisms with varying thickness and width values are shown in Figure 2(a) for the two methods. The error bars indicate the spread of coercive field values with different randomly generated roughness profiles. In all cases a linear decrease in coercive field is seen as the roughness depth increases, with excellent agreement between the two methods. For large roughness levels it is known that the coercive field increases due to strong pinning of domain walls [25] and simulations with large roughness levels, using the rough mesh method, do indeed reproduce this behaviour. For small roughness levels however such pinning effects are not strong enough and the transverse anisotropy introduced by roughness serves mainly to increase the torque from the applied magnetic field, thus lowering the switching, or coercive field.

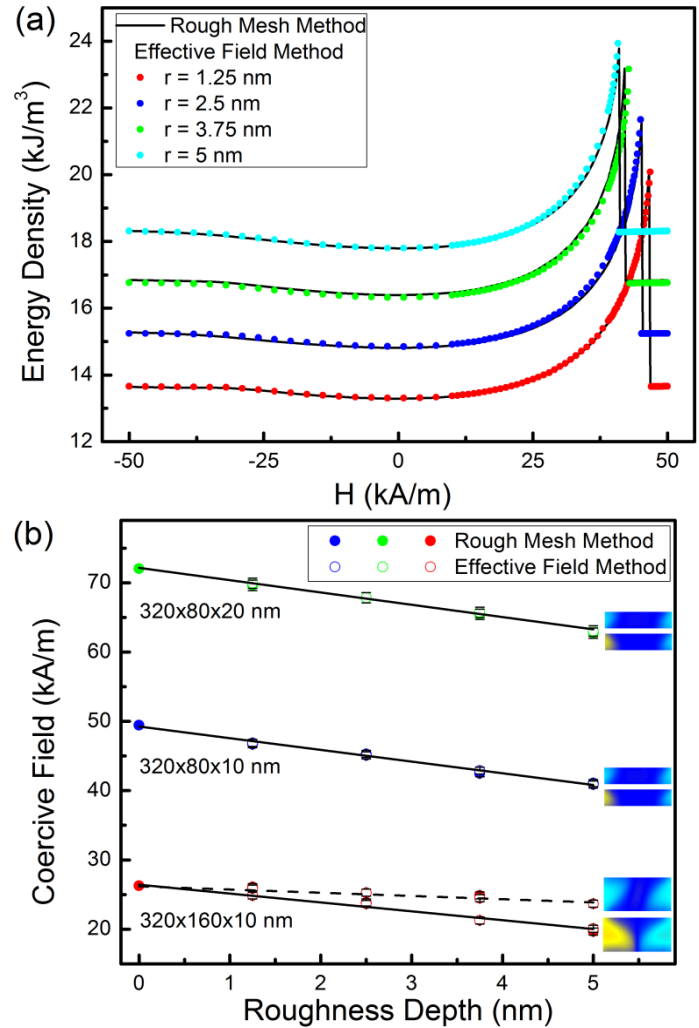

FIG. 2: (color online). Magnetization switching in rectangular prisms with edge roughness calculated using the rough mesh and effective field methods. (a) Energy density (magnetostatic plus exchange energy density) for 4 different roughness depths for a 320×80×10 nm prism and (b) coercive field values for different prism dimensions as a function of roughness depth for the two methods. Inset images show the S-shaped (top) or C-shaped (bottom) magnetization configurations before switching occurs, for each prism.

Depending on the particular roughness profile, the magnetization configuration before switching occurs is found to have either an S-shape or a C-shape [26]. For prisms with large length to width ratio, as for the 80 nm width prisms, the coercive field was not noticeably different between the two modes. For the wider prism however, with 160 nm width, the two switching modes are distinctly separated, as shown in Figure 2(b). In all cases excellent agreement between the two methods was obtained. Further similar tests for a range of different prisms with varying length, width and thickness (not shown here for clarity) have shown an equally good agreement.

## B. Roughness Energy Density Terms

Before continuing with the comparison between the two methods, it is useful to investigate the variation in roughness energy density terms as a function of prism dimensions. In particular for domain wall mobility investigations where a finite calculation region is used, but an effectively infinitely long wire is simulated, it is important to know how long the calculation region should be taken as. Using the standard mesh method the roughness energy density terms were calculated as a function of wire length, width and thickness for a wide range of values of length (40 nm up to 5 μm), width (40 nm up to 640 nm), thickness (1 nm up to 60 nm) and roughness depth (1.25 nm up to 5 nm). The results may be summarised as follows. As the wire length is increased all roughness energy density terms quickly tend to a constant value for all values of width, thickness and roughness depth, as shown in Figure 3 (not all results shown here for clarity). Beyond a length of 1 μm the energy density terms are virtually constant within the normal spread associated with differing randomly generated profiles (indicated by the error bars in Figure 3), thus when considering domain wall mobility calculations it is sufficient to choose a calculation window longer than 1 μm.

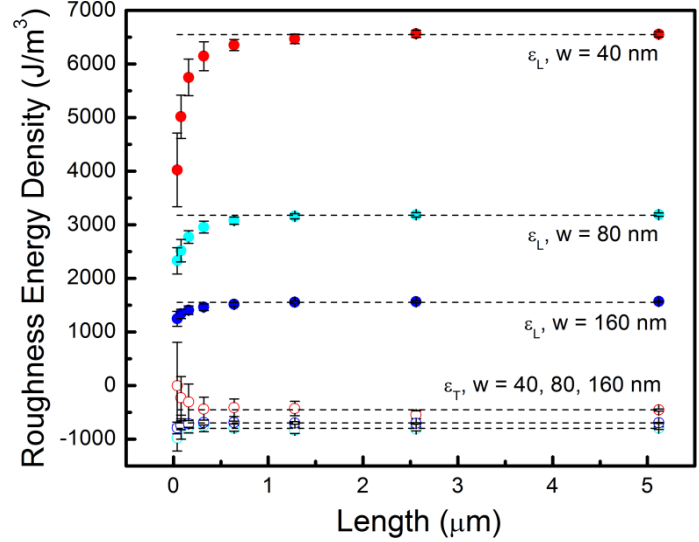

FIG. 3: (color online). Longitudinal (solid circles) and transverse (open circles) roughness energy density terms as a function of length and width for prisms of 20 nm thickness and 2.5 nm edge roughness depth. The error bars indicate the standard deviations obtained from a set of 20 random roughness profiles for each point. The dashed lines are guides for the eye.

The complete dependence of the average roughness energy density terms on dimensions is very complicated – analytical expressions may be obtained using the continuum version of Equation 3, however a few simple rules are worth noting here. All terms are directly proportional to the roughness depth to a good approximation. For edge roughness $<\varepsilon_L>_V$ is always positive and inversely proportional (to a good approximation) to the wire width and, similarly to the length dependence, it quickly tends to a constant value with thickness – above 10 nm thickness $<\varepsilon_L>_V$ is largely constant. On the other hand $<\varepsilon_T>_V$ shows a very complicated dependence with both width and thickness, it is always negative and tends to $<-\varepsilon_L>_V$ for very large values of thickness (above 1 μm, depending on the width).

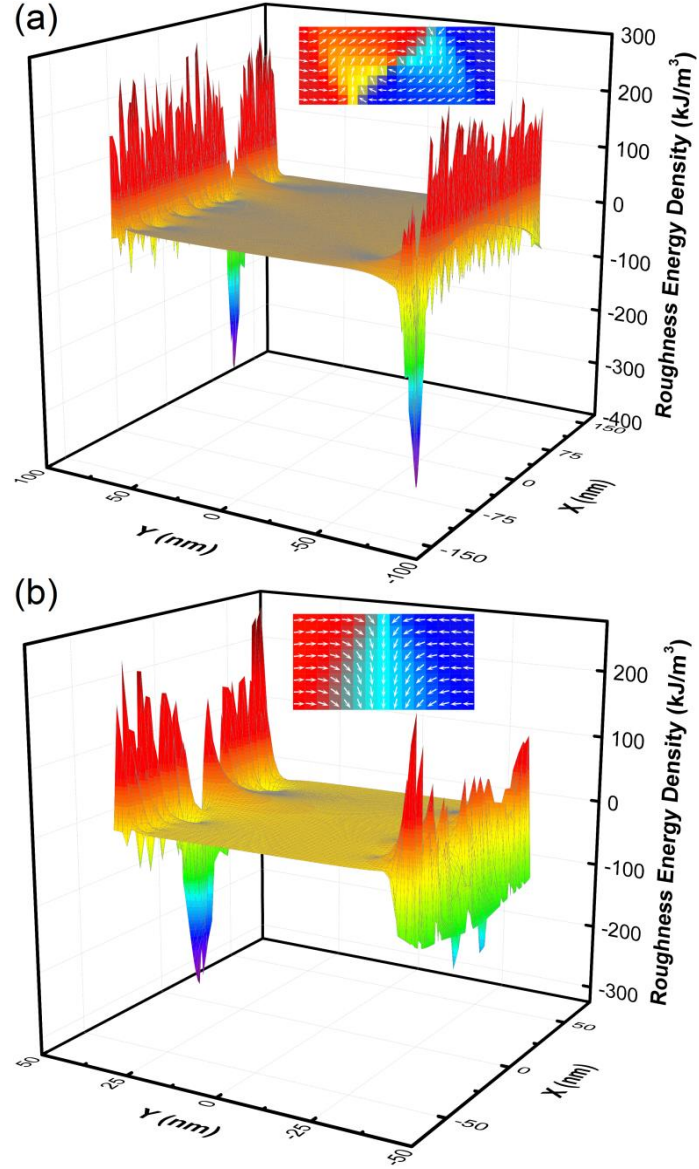

FIG. 4: (color online). Roughness energy density as a function of position using 5 nm edge roughness depth for (a) vortex domain wall and (b) symmetric V-shaped transverse domain wall.

The longitudinal term tends to be highly localized at the edges, however the transverse term has significant contributions even far away from the edges. For surface roughness $<\varepsilon_L>_V$ and $<\varepsilon_T>_V$ are always positive, whilst $<\varepsilon_P>_V$ is always negative and inversely proportional to the thickness (to a good approximation). The longitudinal and transverse terms are localized at the surface, however the perpendicular term has significant contributions throughout the sample volume. As an illustration of this, Figure 4 shows the roughness energy density for vortex and transverse domain walls using a 5 nm edge roughness depth. There are some contributions away from the rough edges, however the largest contributions are at the edges, as expected. Since for edge roughness the easy axis for the roughness configurational anisotropy is transverse to the wire, the energy is in the lowest state for transverse magnetization components, as seen in Figure 4. Thus it should be expected that the movement of a transverse domain wall is strongly affected whilst the movement of a vortex domain wall is less susceptible to edge roughness. Indeed for the latter, surface roughness plays a more important role in thin wires due to the perpendicular magnetization components in the vortex core, as will be discussed in Section IV. The pinning potentials due to roughness tend to be around the same length as the domain wall width at the edges, even though the roughness profile varies quicker than the exchange length. This shows that it is sufficient to include the roughness effective field at the coarse cellsize where the magnetization varies slowly enough for a good continuum approximation.

**C. Domain Wall Mobility – Edge Roughness**

A case that is of particular interest is the use of wire roughness for domain wall mobility calculations. It is well known that wire roughness results in extrinsic pinning of injected domain walls for low driving fields [27], thus it is important to analyse the effects of the effective field roughness model. First, edge roughness is analysed by comparing the two approaches. Field-driven mobility curves calculated using the effective field method with a 5 nm cellsize are shown in Figure 5(a) for an 80 nm wide and 20 nm thick $Ni_{80}Fe_{20}$ wire and edge roughness depth levels ranging from 1.25 nm to 5 nm, containing a symmetric transverse domain wall. As before, tests using a 2.5 nm cellsize showed the same results. The mobility curve calculations start from just below the Walker breakdown field and the field was reduced in steps of 50 A/m. Each field value was applied for 20 ns and the last 10 ns of each step were fitted using linear regression to obtain the domain wall velocity. Typically the velocity changes to a constant value within the first 1 ns and apart from small fluctuations arising from the edge roughness, the domain wall displacement is described very well by a linear dependence on time. The edge roughness profile forms a sequence of pinning potentials which tend to pin the transverse components of domain wall magnetization, since this gives rise to a lower energy configuration. As expected, the pinning field increases with roughness depth, as seen in the inset to Figure 5(a). A time window of 20 ns was chosen since this results in reproducible pinning fields due to the high probability of the domain wall reaching a large pinning potential and becoming trapped; repeated tests did not show any variation in pinning field over this time window, typically the walls become pinned within the first 5 ns after changing the magnetic field. The pinning fields have also been calculated using the rough mesh method and found to be in excellent agreement. The results are shown in the inset to Figure 5(a), where the error bars indicate the discretization of the field step.

In simulating domain wall mobility curves using the rough mesh method, particular care must be taken in choosing the cellsize. It is known that for finite difference meshes, since curved boundaries are discretized using rectangular cells, domain walls can become pinned by the sudden changes in width and also result in drastically reduced mobility [28]. This is a computational artefact and can be reduced either by decreasing the cellsize or by using a correction method, such as the embedded curved boundary method [29]; for a consistent approach to the comparisons the former method was chosen here.

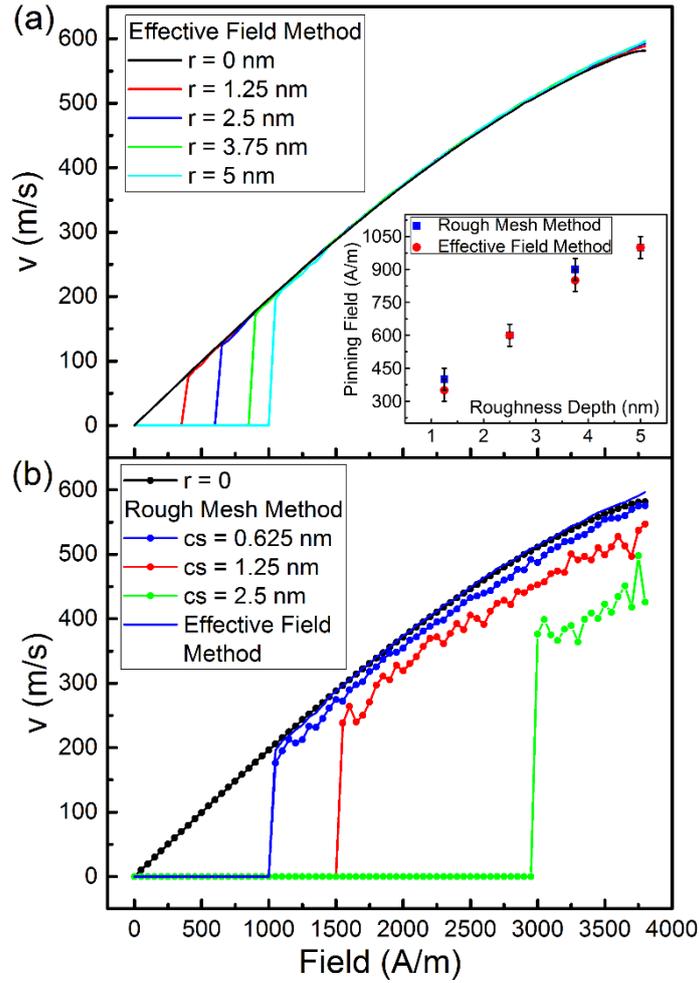

FIG. 5: (color online). Domain wall mobility curves for an 80 nm wide and 20 nm thick wire with varying roughness depth levels calculated using the rough mesh and effective field methods. (a) Effective field method mobility curves. The inset shows the pinning fields obtained using the two methods as a function of roughness depth. The error bars indicate the discretization of the field step. (b) Domain wall mobility curves obtained using the rough mesh method with 5 nm roughness depth and different cellsize values, compared to the effective field method for the same roughness depth.

Mobility curves for the 5 nm roughness depth are shown in Figure 5(b) for cellsize values ranging from 2.5 nm down to 0.625 nm. A good match was obtained for the 0.625 nm cellsize, with the larger cellsize values being clearly inadequate to accurately calculate the domain wall mobility. The pinning field values in the inset to Figure 5(a) were consequently calculated using a 0.625 nm cellsize. Simulations with a smaller cellsize are impractical partly due to the increased problem size but very significantly due to the stiffness of the LLG equation requiring very small time steps (~ 1 fs for a fourth order explicit scheme); using an implicit evaluation scheme did not improve the computation time.

## IV. SURFACE ROUGHNESS

Finally, the effective field roughness model is applied to vortex domain wall mobility calculations in wires with surface roughness. Surface roughness, as well as magnetic defects, is known to result in pinning effects on moving vortex domain walls. [30,31] Here a 320 nm wide and 40 nm thick $Ni_{80}Fe_{20}$ wire is investigated, containing a vortex domain wall. The domain wall mobility is calculated, using 3D simulations, as a function of surface roughness depth by increasing the field in steps of 50 A/m up to the Walker breakdown threshold, which was found to be 1000 A/m in this case. A roughness energy density plot obtained using both the rough mesh method and effective field method using Equation 5 as before, is shown in Figure 6(a) for a surface roughness depth of 2.5 nm. In this case the roughness contributes an effective anisotropy with easy axis perpendicular to the surface, thus providing pinning potentials for the perpendicular components of magnetization, most significantly for the vortex core which has magnetization components pointing out-of-plane. Vortex domain wall mobility curves are shown in Figure 6(b). Above a roughness depth of 1.5 nm the vortex wall can be pinned and three regimes can be distinguished: i) uniform translation at low fields, ii) dynamic pinning and stochastic creep regime and iii) depinning and uniform translation.

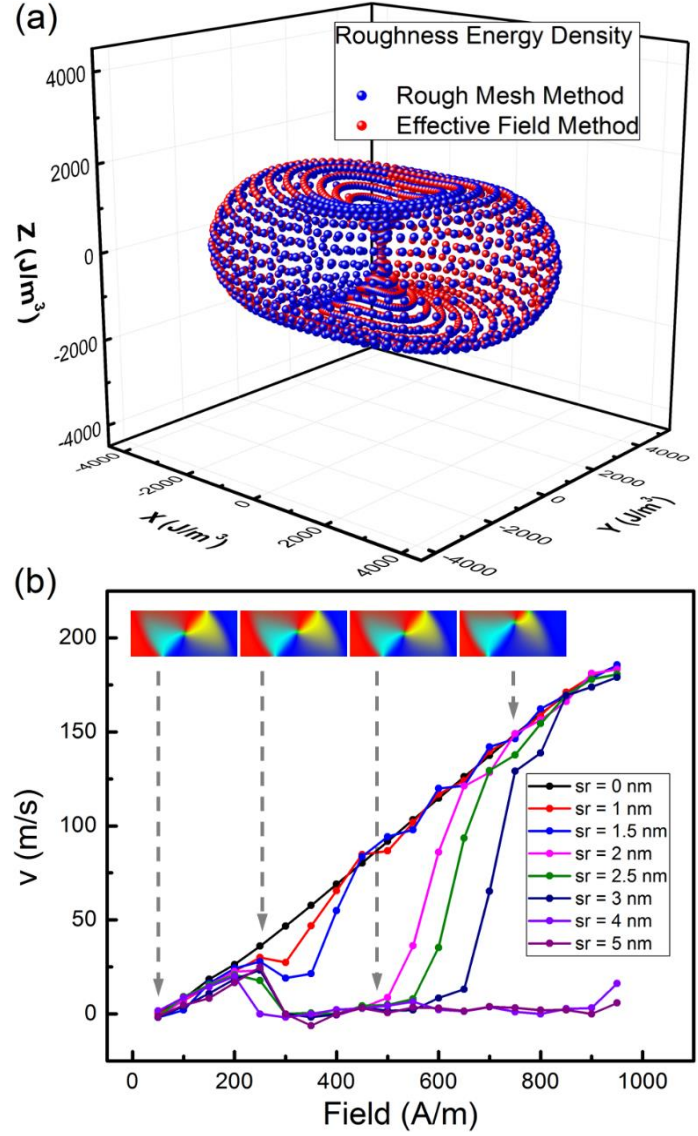

FIG. 6: (color online). (a) Surface roughness energy density calculated for a uniformly magnetized 320 nm wide and 40 nm thick wire with 2.5 nm roughness depth, with the lowest energy point shifted to zero. (b) Domain wall mobility curves for a vortex domain wall in this wire with varying levels of surface roughness. The insets shows the vortex magnetization configuration at important points along the mobility curve for 3 nm surface roughness depth.

For surface roughness depths above 3 nm the wall remains pinned all the way to the 1000 A/m threshold. As expected, increasing the surface roughness depth results in stronger pinning and in this regime only very small and stochastic domain wall creep is observed – the vortex core can become unpinned but is then quickly pinned again by neighbouring pinning potentials due to surface roughness; this results in the small variations in velocity in this regime as seen in Figure 6(a). The effects of thermal excitations have not been studied here as the main purpose of this work was to introduce and analyse the effective field roughness model; the domain wall movement stochasticity arises solely due to the surface roughness – slight oscillations in the magnetization components, under the applied magnetic field, cause the vortex core to jump small distances between the random pinning sites, resulting in an average creep velocity of ~ 2 m/s. What is more surprising is that at low fields the wall translates uniformly almost independently of surface roughness depth. As the field is increased the vortex core gradually drifts towards one of the wire edges and at a critical field, between 250 and 300 A/m, the vortex core becomes pinned and the uniform translation stops. As the vortex core becomes pinned, it is observed to relax back to the wire centre and the entire vortex configuration is slightly shrunk as compared to the low field uniform translation mode. This is reflected by a steep increase in the average roughness energy density just before the vortex core becomes pinned – since the roughness energy density is larger for longitudinal magnetization components, see Figure 6(a), shrinking of the vortex structure results in a greater contribution from the longitudinal components. The roughness energy density as a function of time during the pinning event is shown in Figure 7. During the uniform translation mode the vortex spin structure is known to oscillate [32,33]. This is also observed in micromagnetic simulations and results in the oscillation in roughness energy density seen in Figure 7 before the start of the pinning event. As the vortex structure becomes distorted with increasing magnetic field, the interaction between the vortex core and surface roughness supresses this oscillation, as seen in Figure 7, which forms the onset of the pinning event. This process is also illustrated by the insets in Figure 6(b), although the changes in magnetization structure described are small. After the depinning field is reached, the vortex core quickly jumps close to one of the wire edges, following which the wall moves uniformly. Further increasing the field causes the vortex core to drift closer and closer to the wire edge until Walker breakdown occurs at 1000 A/m in all cases (apart from the strongly pinned 4 and 5 nm roughness depth cases).

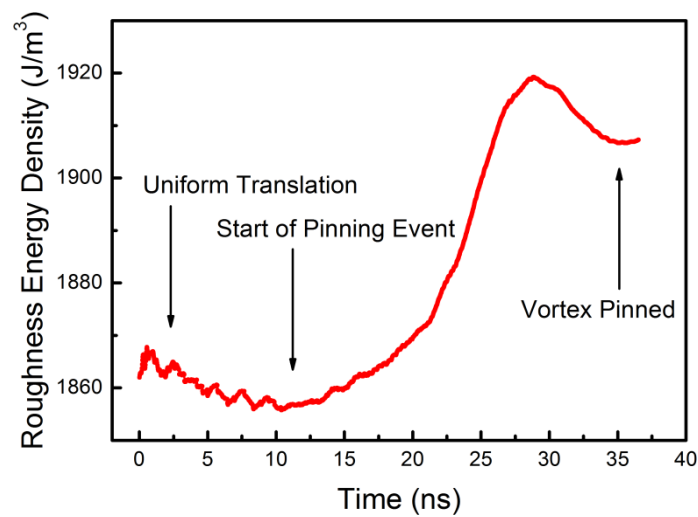

FIG. 7: (color online). Average roughness energy density as a function of time, showing a vortex domain wall pinning event.

# V. SUMMARY

An effective field model of small roughness levels in magnetic structures was introduced as a perturbation on the magnetostatic energy of a corresponding smooth structure. The model is generally applicable to any type of roughness profile and to both edge and surface roughness. Small roughness levels, below the exchange length, have been shown to result in a configurational anisotropy and the resulting effective roughness fields are sufficient to describe the effects of roughness on magnetization structures without the explicit need to refine the computational cellsize beyond what is normally required for the corresponding smooth structure. The model was validated using a series of tests, including magnetization switching and domain wall mobility calculations, against the standard rough mesh method which uses a much smaller computational cellsize in order to accommodate the roughness profile directly. For small edge roughness levels, the coercive field in magnetization switching simulations was found to decrease linearly. This is due to the increase of the torque from the applied magnetic field on transverse magnetization components, which have a lower edge roughness energy density. Domain wall mobility calculations for transverse domain walls have shown that edge roughness results in extrinsic pinning, increasing in strength with roughness depth, in agreement with the standard rough mesh method. Vortex domain walls have been shown to be highly susceptible to surface roughness, resulting in dynamic pinning of vortex cores. Three vortex domain wall movement regimes have been found, uniform translation at low fields independent of surface roughness, vortex pinning regime with stochastic domain wall creep and depinning followed by uniform translation at higher fields up to the breakdown threshold.

**Appendix A: Methods**

All simulations were done using the micromagnetics software Boris [34] written by the author. The software was fully tested against standard micromagnetics problems. The Landau-Lifshitz-Gilbert (LLG) equation was solved using a finite difference mesh. A number of evaluation methods were used, $2^{nd}$ order Adams-Bashforth-Moulton (ABM2) predictor-corrector scheme with quadratic interpolation on time-step change, Runge-Kutta-Fehlberg (RKF45) adaptive time-step and $4^{th}$ order Runge-Kutta (RK4) fixed time-step. ABM2 was found to be slightly more computationally efficient than RKF45 for domain wall mobility simulations whilst RKF45 was more efficient in magnetization switching simulations, mainly due to its more stable time-step across the larger range of magnetic fields. For simulations with cellsize of 0.625 nm neither ABM2 nor RKF45 were suitable and RK4 was used. An implicit scheme using the $2^{nd}$ order backward differentiation formula (BDF2) with direct Newton-Raphson solver was also tested, however no computational advantage was found compared to the explicit scheme. The magnetostatic term was computed using FFT-based convolution with Radix-4 FFT (Radix-2 step for odd powers of 2); the lower arithmetic operations count split-radix FFT was found to be less efficient due to its greater number and less cache-friendly memory access instructions. All FFT routines, and other critical computation routines, used here were written directly in assembler using the SIMD AVX instruction set, resulting in a speed-up factor of around 4 compared to GCC or MSVC compiler-generated routines. For the larger 3D simulations CUDA-based Radix-4 FFTs were used; all computations were performed in double floating-point precision. The exchange term was computed using the 6-neighbor scheme with Neumann boundary conditions. For domain wall mobility calculations a moving mesh algorithm was used with spin-wave absorbing boundaries at both ends. To simulate an effectively infinite wire length, uniform magnetization continuations of the wire were set at both ends outside of the mesh and the resulting magnetic field inside the mesh was calculated.

Standard values for $Ni_{80}Fe_{20}$ were used, namely $M_s = 8\times10^5$ (A/m), $A = 1.3\times10^{-11}$ (J/m) and $\alpha = 0.02$. For the effective field method calculations, a cellsize of 5 nm was used as this was sufficiently fine to accurately reproduce changes in magnetization – test simulations using a smaller cellsize of 2.5 nm did not show any significant differences. For the mesh method the cellsize varied between 0.625 nm and 5 nm as detailed in the main text. Granular roughness profiles were used with the mean grain diameter equal to the roughness depth. Langevin dynamics have not been considered here as the main purpose of this work was to introduce and analyse the effective field roughness model, however it is hoped this work will stimulate further investigations in this area.

[1] K.J.A. Franke, B. Van de Wiele, Y. Shirahata, S.J. Hämäläinen, T. Taniyama and S. van Dijken, "Reversible electric-field-driven magnetic domain-wall motion" Physical Review X **5**, 011010 (2015).

[2] S. Lepadatu, M. C. Hickey, A. Potenza, H. Marchetto, T. R. Charlton, S. Langridge, S. S. Dhesi, and C. H. Marrows, "Experimental determination of spin-transfer torque nonadiabaticity parameter and spin polarization in permalloy," Phys. Rev. B 79(9), 094402 (2009).

[3] L. San Emeterio Alvarez, K.-Y. Wang, S. Lepadatu, S. Landi, S.J. Bending, and C.H. Marrows, "Spin-Transfer-Torque-Assisted Domain-Wall Creep in a Co-Pt Multilayer Wire", Phys. Rev. Lett. **104**, 137205 (2010).

[4] A.T. Galkiewicz, L. O'Brien, P.S. Keatley, R.P. Cowburn and P.A. Crowell, "Resonance in magnetostatically coupled transverse domain walls" Phys. Rev. B **90**, 024420 (2014).

[5] E.Martinez, "The stochastic nature of the domain wallmotion along high perpendicular anisotropy strips with surface roughness," Journal of Physics **24**, 024206 (2012).

[6] Y. Nakatani. A. Thiaville and J. Miltat, "Faster magnetic walls in rough wires", Nature Materials **2**, 512 (2003).

[7] A. Thiaville, Y. Nakatani, J. Miltat and Y. Suzuki, "Micromagnetic understanding of current-driven domain wall motion patterned nanowires", Europhys. Lett. **69**, 990 (2005).

[8] M. Albert, M. Franchin, T. Fischbacher, G. Meier and H. Fangohr, "Domain wall motion in perpendicular anisotropy nanowires with edge roughness", J. Phys. Cond. Mat. **24**, 024219 (2012).

[9] H. Min, R.D. McMichael, M.J. Donahue, J. Miltat and M.D. Stiles, "Effects of disorder and internal dynamics on vortex wall propagation", Phys. Rev. Lett. **104**, 217201 (2010).

[10] J. Leliaert, B. Van de Wiele, A. Vansteenkiste, L. Laurson, G. Durin, L. Dupré and B. Van Waeyenberge, "A numerical approach to incorporate intrinsic material defects in micromagnetic simulations" J. Appl. Phys. **115**, 17D102 (2014).

[11] G. Tatara, T. Takayama, H. Kohno, J. Shibata, Y. Nakatani, H. Fukuyama, "Threshold current of domain wall motion under extrinsic pinning, β-term and non-adiabaticity" J. Phys. Soc. Jpn. **75**, 064708 (2006).

[12] H. Saarikoski, H. Kohno, C.H. Marrows and G. Tatara, "Current-driven dynamics of coupled domain walls in a synthetic antiferromagnet" Phys. Rev. B **90**, 094411 (2014).

[13] P. Sinha, N.A. Porter and C.H. Marrows, "Strain-induced effects on the magnetic and electronic properties of epitaxial $Fe_{1-x}Co_xSi$ thin films" Phys. Rev. B **89**, 134426 (2014).

[14] H. Kaiju, N. Basheer, K. Kondo and A. Ishibashi, "Surface roughness and magnetic properties of Ni and $Ni_{78}Fe_{22}$ thin films on polyethylene naphthalate organic substrates" IEEE Trans. Mag. **46**, 1356 (2010).

[15] P. Bruno, "Magnetic surface anisotropy of cobalt and surface roughness effects within Néel's model" J. Phys. F: Met. Phys. **18**, 1291 (1988).

[16] P. Bruno, "Dipolar magnetic surface anisotropy in ferromagnetic thin films with interfacial roughness" J. Appl. Phys. **64**, 3153 (1988).

[17] C.A.F. Vaz, S.J. Steinmuller and J.A.C. Bland "Roughness-induced variation of magnetic anisotropy in ultrathin epitaxial films: The undulating limit" Phys. Rev. B **75**, 132402 (2007).

[18] M.J. Donahue and R.D. McMichael, "Micromagnetics on curved geometries using rectangular cells: error correction and analysis" IEEE Trans. Mag. **43**, 2878 (2007).

[19] W.F. Brown, Jr. "Magnetostatic Principles in Ferromagnetism" Amsterdam, The Netherlands: North-Holland (1962).

[20] A.J. Newell, W. Williams and D.J. Dunlop, "A generalization of the demagnetizing tensor for nonuniform magnetization" J. Geophysical Research-Solid Earth **98**, 9551 (1993).

[21] The meaning of small roughness level is precisely the range of validity of the roughness model – here roughness levels up to the exchange length were considered as this is the range of roughness values over which the model is most useful.

[22] Equation 4 is calculated on a fine sub-mesh using FFT-based convolution by separating it into a sum of two terms, and the energy is averaged for each cell in the coarse mesh.

[23] G.S. Abo, Y.-K. Hong, J. Park, J. Lee, W. Lee and B-C. Choi, "Definition of magnetic exchange length" IEEE Trans. Mag. **49**, 4937 (2013).

[24] M.J. Donahue, "Micromagnetic investigation of periodic cross-tie/vortex wall geometry" Adv. Cond. Mat. Phys. Article ID 908692 (2012).

[25] M.T. Bryan, D. Atkinson and R.P. Cowburn, "Edge roughness and coercivity in magnetic nanostructures" Journal of Physics **17**, 40 (2005).

[26] X.X. Liu, J.N. Chapman, S. McVitie and C.D.W. Wilkinson, "Introduction and control of metastable states in elliptical and rectangular magnetic nanoelements" Appl. Phys. Lett. **84**, 4406 (2004).

[27] G.S.D. Beach, C. Nistor, C. Knutson, M. Tsoi and J.L. Erskine, "Dynamics of field-driven domain-wall propagation in ferromagnetic nanowires" Nature Materials **4**, 741 (2005).

[28] H. Riahi, F. Montaigne, N. Rougemaille, B. Canais, D. Lacour and M. Hehn, "Energy levels of interacting curved nano-magnets in a frustrated geometry : increasing accuracy when using finite difference methods" J. Phys.: Cond. Mat. **25**, 296001 (2013).

[29] M.R. Gibbons, G. Parker, C. Cerjan and D.W. Hewett, "Finite difference micromagnetic simulation with self-consistent currents and smooth surfaces" Physica B **275,** 11 (1999).

[30] T.Y. Chen, M.J. Erickson, P.A. Crowell and C. Leighton, "Surface roughness dominated pinning mechanism of magnetic vortices in soft ferromagnetic films" Phys. Rev. Lett. **109**, 097202 (2012).

[31] J. Leliaert, B. Van de Wiele, A. Vansteenkiste, L. Lauson, G. Durin, L. Dupré and B. Van Waeyenberge, "Influence of material defects on current-driven vortex domain wall mobility", Phys. Rev. B **89**, 064419 (2014).

[32] A. Bisig et al., "Correlation between spin structure oscillations and domain wall velocities" Nature Communications **4**, 2328 (2013).

[33] A. Vansteenkiste, K.W. Chou, M. Weigand, M. Curcic, V. Sackmann, H. Stoll, T. Tyliszczak, G. Woltersdorf, C.H. Back, G. Schütz and B. Van Waeyenberge, "X-ray imaging of the dynamic magnetic vortex core deformation" Nature Physics **5**, 332 (2009).

[34] M.M. Vopson, S. Lepadatu, "Solving the electrical control of magnetic coercive field paradox", Appl. Phys. Lett. **105**, 122901 (2014).